\documentclass{ws-ijmpa}
\newcommand{\adb}{\allowdisplaybreaks }
\newcommand{\TE}{{\rm TE}}
\newcommand{\TM}{{\rm TM}}
\newcommand{\te}{\textsf{\scriptsize TE} }
\newcommand{\tm}{\textsf{\scriptsize TM} }
\newcommand{\bs}{\begin{subequations}}
\newcommand{\es}{\end{subequations}}

\begin{document}

\markboth{Nail Khusnutdinov}
{The Casimir-Polder interaction an atom with spherical shell}

\catchline{}{}{}{}{}
%
%%%%%%%%%%%%%%%%%%%%%%%%%%%%%%%%%%%%%%%%%%%%%%%%%%%%%%%%%%%%%%%%%%%%

\title{THE CASIMIR-POLDER INTERACTION AN ATOM WITH SPHERICAL SHELL}

\author{NAIL KHUSNUTDINOV}

\address{Institute of Physics, Kazan Federal University, Kremlevskaya
18\\
Kazan, 420008, Russia\\
7nail7@gmail.com}

\maketitle

\begin{history}
\received{23 June 2011}
%\revised{Day Month Year}
\end{history}

\begin{abstract}
The Casimir-Polder and van der Waals interaction energy of an atom with
infinitely thin sphere with finite conductivity is investigated in the framework
of the hydrodynamic approach. We put the sphere into spherical cavity inside the
infinite dielectric media, then calculate the energy of vacuum fluctuations in
the context of the zeta-function approach. The energy for a single atom is
obtained by rarefying media. The Casimir-Polder expression for an atom and plate
is recovered in the limit of the infinite radius of the sphere. Assuming a
finite radius of the sphere, the interaction energy of an atom falls down
monotonic as third power of distance between atom and sphere for short distance
and as seventh power for large distance from the sphere.

\end{abstract}

\ccode{PACS numbers: 73.22.-f, 34.50.Dy, 12.20.Ds}

\section{Introduction}\label{Sec:Intro}
The general theory of the van der Waals force was developed by Lifshits in Refs.
\refcite{Lif56,LifPit80} in the framework of  statistical physics. In the case
of
interaction between particle and plate it is commonly referred to as the
Casimir-Polder force.\cite{CasPol48} For small distance the potential of
interaction is proportional to inverse third degree of distance from the plate.
For large distance the retardation of the interaction is taken into account and
the potential falls down as fourth degree of distance. The last achievements in
Casimir effect have been discussed in great depth in books and
reviews.\cite{Mil01}\cdash\cite{KliMohMos09}

The van der Waals force is very important for interaction of graphene
(graphite layers) with
bodies\cite{BogOveHylLunBruJen00}\cdash\cite{BorFiaGitVas09} and
microparticles.\cite{BonLam05}\cdash\cite{ChuFedKliYur10}
An understanding of the mechanisms of molecule-nanostructure interaction is of
importance for the problem of hydrogen storage in carbon
nanostructures.\cite{DilJonBekKiaBetHeb97} The microscopic mechanisms underlying the
absorption phenomenon remain unclear (see, for example
Ref.~\refcite{Nec06}).

In the present paper we use model of the fullerene in terms of the two
dimensional free electron gas\cite{Fet73} which is usually called as
hydrodynamical model.  This model was applied and developed for the molecule
$C_{60}$ in Ref.~\refcite{Bar04}, for flat plasma sheet  in Ref.
\refcite{BorPirNes05} and for spherical plasma surface in Ref.
\refcite{BorKhu08}. In
the framework of this model the conductive surface is considered as infinitely
thin shell with the specific wave number $\Omega = 4\pi n e^2/mc^2$, where $n$
is surface density of electrons and $m$ is the electron mass. Since the surface
is infinitely thin, the information about the properties
of the surface is encoded in the boundary conditions on the conductive surface
which are different for TE and TM modes. It was
shown\cite{BorKhu08} that
the energy of the vacuum electromagnetic fluctuations for surface shaped as
sphere has a maximum for radius of sphere approximately equal to the specific
wavelength of the model $\lambda_\Omega = 2\pi/\Omega$. What this means is the
Casimir force tries to enlarge sphere with radius larger then $\lambda_\Omega$
and it tries to reduce the sphere with radius larger then $\lambda_\Omega$. The
Boyer result\cite{Boy68} is recovered in the limit $\Omega \to \infty$.

At the same time it is well known\cite{GeiNov07} that the energy of electrons
in graphene has linear frequency dependence whereas in framework of the
hydrodynamic model the energy of electrons is quadratic in the frequency. There
is also another point that the electrons in the graphene have zero or very small
effective mass. To describe correctly these unusual properties of electrons in
graphene the Dirac fermion model was suggested in Ref. \refcite{Sem84}. The
electrons in this model are described by $(2+1)D$ Dirac action with
characteristic propagation velocity as Fermi velocity $v_F \approx c/300$ and
very small mass gap $m < 0.1 eV$. This model was applied for calculation of
Casimir interaction energy between graphene plate and perfect conductor
plane\cite{BorFiaGitVas09} and recently
for Casimir-Polder interaction energy between graphene and H, He$^*$ and Na
atoms.\cite{ChuFedKliYur10}

It was shown that the Casimir energy for large distance between graphene plate
and perfect conductor plane\cite{BorFiaGitVas09} is decreasing by one power of
the separation a faster than for ideal conductors, that is as $(ma)^{-4}$. If
the mass of gap is zero at the beginning of calculations, $m=0$, they obtained
standard dependence $a^{-3}$. For the case  of Casimir-Polder interaction
energy between graphene and atoms\cite{ChuFedKliYur10} the hydrodynamic and
the Dirac models give qualitatively different results. For the large separation
the energy decreases with separation as $a^{-4}$ which is a typical behavior of
the atom-plate interaction at relativistic separations, but the coefficients are
different. In the case of H, He$^*$ and Na atoms, the hydrodynamic model gives
$\approx 5$ times larger coefficient than the Dirac model. There is also
interesting observation about mass gap parameter: the energy does not depend on
the parameter for $m < 10^{-3/2} eV$ and therefore the limit $m\to 0$ is
satisfied.

In the present paper the hydrodynamical model of fullerene is adopted -- the
infinitely thin sphere with radius $R$ in vacuum and finite conductivity. To
obtain the van der Waals interaction energy between an atom and this sphere we
use the following approach which is due to Lifshits (see Refs.
\refcite{Lif56,LifPit80,BorGeyKliMos06,BlaKliMos07}). We put the sphere inside
the spherical vacuum cavity with radius $L = R+d$ which is inside the
dielectric media with coefficients $\mu, \varepsilon$. Then we find the
zero-point energy of this system by using the zeta-function regularization
approach, and take the limit of the rared media with $\varepsilon = 1+ 4\pi N
\alpha + O(N^2)$, where $N\to 0 $
is the volume density of the atoms and $\alpha$ is the polarizability of the
unit atom. The interaction energy per unit atom which is situated $d$ from the
sphere is found by simple formula
\begin{equation*}
E_a(s) = -\lim_{N\to 0}\frac{\partial_d E(s)}{4\pi N (R+d)^2},
\end{equation*}
where $E(s)$ is the zeta-regularized energy with regularization parameter $s$.

\section{Matching Conditions For Two Cylinders}\label{Sec:Maxwell}

Let us consider a conductive infinitely thin sphere with radius $R$ in vacuum
spherical cavity with radius $L=R+d$ which is inside the dielectric media with
parameters $\mu, \varepsilon$. We have two concentric spheres and we should
consider the boundary conditions on two spherical boundaries.

Assuming the spherical symmetry, the electromagnetic field is factorized for two
independent polarizations usually called as \te and \tm modes. The angular
dependence is described by spherical functions $Y_{lm}$ and radial function $f$
subjects for radial equation
\begin{equation}
f'' + \frac 2r f' + \left( \frac{\omega^2}{c^2}\varepsilon \mu  -
\frac{l(l+1)}{r^2}\right)f = 0.
\end{equation}
The two independent solutions of this equation are the spherical Bessel
functions $j_l(z) = \sqrt{\pi/2z} J_{l+1/2}(z), \ y_l(z) = \sqrt{\pi/2z}
Y_{l+1/2}(z)$, where $z = r \omega \sqrt{\varepsilon\mu}/c$.

At the boundary, $L=R+d$, the matching conditions read
\bs\label{matching_cond}
\begin{eqnarray}
 \mathbf{n} \cdot [\mathbf{B}_2 - \mathbf{B}_1]_L &=& 0, \ \mathbf{n} \cdot
[\mathbf{D}_2 - \mathbf{D}_1]_L =0,\adb\\
 \mathbf{n} \times [\mathbf{H}_2 - \mathbf{H}_1]_L &=& 0, \ \mathbf{n} \times
[\mathbf{E}_2 - \mathbf{E}_1]_L =0,
\end{eqnarray}
\es
where $\mathbf{n} = \mathbf{r}/r$ is an unit normal to the sphere. We have to
take into account also that $k=\omega /c$ inside the sphere $r=L$ and $k= \omega
\sqrt{\mu\varepsilon}/c$ outside the sphere. The square brackets above denote
the coincidence limit on the boundary $r=L$.

The electromagnetic fields given infinitely thin conductive surface $\Sigma$
in vacuum was considered by Fetter.\cite{Fet73} The applications of
this model for vacuum fluctuations of field see in Refs.
\refcite{Bar04}--\refcite{BorKhu08}. The  boundary conditions on the
sphere with $r=R$ read
\bs \label{cond2}
\begin{eqnarray}
\mathbf{n}\cdot [\mathbf{H}_{2} - \mathbf{H}_{1}]_R &=& 0,\ \mathbf{n}\cdot
[\mathbf{E}_{2} - \mathbf{E}_{1}]_R = \frac{\Omega }{k^2} \nabla_{|\!|} \cdot
\mathbf{E}_{|\!|}, \\
\mathbf{n} \times [\mathbf{H}_2 - \mathbf{H}_1]_R &=& -\frac{i\Omega }{k }
\mathbf{n} \times \mathbf{E}_{|\!|},\ \mathbf{n} \times [\mathbf{E}_2 -
\mathbf{E}_1]_R = 0,
\end{eqnarray}
\es
where $k=\omega /c$ and $\Omega = 4\pi n e^2/mc^2$ is a specific wave number on
the sphere. Because of the fact that the sphere is infinitely thin we may
consider the Maxwell equations in vacuum with zero right hand side
and all information about sphere will be encoded in boundary conditions
(\ref{cond2}). An interesting treatment of this boundary condition is in Ref.
\refcite{Vas09}.

\section{The Solution of the Matching Conditions}\label{Sec:Matching}

Let us represent the radial function in the following way
\begin{equation}
f = \left\{
 \begin{array}{lll} f_{in}&=a_{in} j_l(kr),& r<R \\
               f_{out}&=a_{out} j_l(kr) + b_{out} y_{l}(kr),&R<r<L \\
               f_{\varepsilon}&=a_{\varepsilon} h^{(1)}_l(kr),&r>L
 \end{array}\right.
\end{equation}
where $j_l,y_l$ and $h_l^{(1)}$ are the spherical Bessel functions and $k=\omega
/c$ inside the sphere, $r < L$ and $k= \omega \sqrt{\mu\varepsilon}/c$ outside
the sphere for $r > L$.

In this case the matching conditions (\ref{matching_cond}) and (\ref{cond2}) in
manifest form read
\begin{eqnarray}
 [rf_{out} - rf_{in}]_{R} &=&0,\nonumber\adb\\{}
[(rf_{out})'_r -  (rf_{in})'_r - \Omega (rf_{in})]_R &=&0,\nonumber\adb\\{}
[rf_{out} - rf_{\varepsilon}]_{L} &=&0,\adb\\{}
[(rf_{out})'_r -  \frac{1}{\mu}(rf_{\varepsilon})'_r]_L &=&0, \nonumber
\end{eqnarray}
for \te mode, and
\begin{eqnarray}
 [(rf_{out})'_r - (rf_{in})'_r]_{R} &=&0,\nonumber\adb\\{}
[(rf_{out}) -  (rf_{in}) + \frac{\Omega}{k^2} (rf_{in})'_r]_R
&=&0,\nonumber\adb\\{}
[rf_{out} -  \frac{1}{\mu}rf_{\varepsilon}]_L&=&0,\adb\\{}
[(rf_{out})'_r -  \frac{1}{\mu\varepsilon}(rf_{\varepsilon})'_r]_L
&=&0,\nonumber
\end{eqnarray}
for \tm mode. The solutions of these equations exist if and only if the
following equations are satisfied ($\mu=1$)
\bs
\begin{eqnarray}
 \Sigma_{\te} &=& H'(z_\varepsilon) \Psi_{\te} -
\frac{1}{\sqrt{\varepsilon}} H(z_\varepsilon) \Psi'_{\te}=0 ,\adb\\
 \Sigma_{\tm} &=& z^2\left\{ H(z_\varepsilon) \Psi'_{TM} -
\frac{1}{\sqrt{\varepsilon}} H'(z_\varepsilon) \Psi_{\tm}\right\}=0 ,
\end{eqnarray}
\es
where $z_\varepsilon = z \sqrt{\mu\varepsilon}$, $z = kL = \omega L/c$; the
prime is derivative with respect the argument, and
\bs
\begin{eqnarray}
 \Psi_{\te}(z) &=& J(z) + \frac{\Omega}{k} J(x) [J(x) Y(z) - J(z) Y(x)],\adb\\
 \Psi_{\tm}(z) &=& J(z) + \frac{\Omega}{k} J'(x) [J'(x) Y(z) - J(z) Y'(x)].
\end{eqnarray}
\es
Here $J(x) = xj_l(x),\ Y(x) = xy_l(x),\ H(x) = xh^{(1)}_l(x)$ are the
Riccati-Bessel functions, and $x=kR$. For $\varepsilon =1$, the result obtained
in the Ref. \refcite{BorKhu08} is
recovered
\bs
\begin{eqnarray}
 \Sigma_{\te} &=& i\left\{1 - \frac{\Omega}{ik} J(x)H(x)
\right\}=if_{\te}(k),\adb\\
 \Sigma_{\tm} &=& -iz^2 \left\{1 - \frac{\Omega}{ik} J'(x)H'(x) \right\} = -
iz^2f_{\tm}(k),
\end{eqnarray}
\es
for real value of $k$.

On the imaginary axis $k\to i k$ we obtain
\bs \label{sigmas}
\begin{eqnarray}
 \Sigma_{\te} &=& \frac{1}{\sqrt{\varepsilon}} e_l(z_\varepsilon)
\Phi'_{\te} - e'_l(z_\varepsilon) \Phi_{\te}  ,\adb\\
 \Sigma_{\tm} &=& z^2\left\{e_l(z_\varepsilon) \Phi'_{\tm} -
\frac{1}{\sqrt{\varepsilon}} e_l'(z_\varepsilon) \Phi_{\tm}\right\},\adb\\
 \Phi_{\te} &=& s_l(z) + \frac{Q}{x} s_l(x) [s_l(z) e_l(x) - s_l(x)
e_l(z)],\adb\\
 \Phi_{\tm} &=& s_l(z) - \frac{Q}{x} s'_l(x) [s_l(z) e'_l(x) - s'_l(x) e_l(z)],
\end{eqnarray}
\es
where $Q=\Omega R$, $z = kL,\ z_\varepsilon = z \sqrt{\varepsilon},\ x = kR$,
$\varepsilon = \varepsilon (i\omega)$ and
\begin{equation}
 s_l(x) = \sqrt{\frac{\pi x}{2}} I_{l+1/2}(x), \ e_l(x) = \sqrt{\frac{2 x}{\pi}}
K_{l+1/2}(x)
\end{equation}
are the Riccatti-Bessel spherical functions of the second kind. For $\varepsilon
= 1$ we obtain
\begin{equation}
 \Sigma_{\te} = f_{\te}(ik), \ \Sigma_{\tm} = z^2f_{\tm}(ik)
\end{equation}
in accordance with Ref. \refcite{BorKhu08}.

\section{The Energy}\label{Sec:Energy}
Within the limits of approach,\cite{BorEliKirLes97} the
expressions for \te and \tm contributions in regularized  zero-point energy read
($\omega=kc,\nu =  l + 1/2$)
\begin{eqnarray}
E^{\te}(s) &=& -\frac{\hbar c\cos\pi s}{\pi} \mu^{2s}\sum_{l=1}^\infty \nu
\int_0^\infty dk k^{1-2s} \partial_k \ln \Sigma_{\te},\adb\\
E^{\tm}(s) &=& -\frac{\hbar c\cos\pi s}{\pi} \mu^{2s} \sum_{l=1}^\infty \nu
\int_0^\infty dk k^{1-2s} \partial_k \ln  \Sigma_{\tm},
\end{eqnarray}
where the integrand functions are given by Eqs. (\ref{sigmas}).

Let us consider now the rared media with $\varepsilon (i\omega) = 1 + 4\pi N
\alpha (i\omega) + O(N^2)$, where $\alpha$ is polarizability of the atom and the
density of the dielectric matter $N\to 0$. In this case the Casimir energy
$E(s)$ is expressed in terms the energy per unit atom $E_a(s)$ by relation
\begin{equation}
E(s) = N \int_d^\infty E_a(s) 4\pi (R+r)^2 dr + O(N^2).
\end{equation}
From this expression it follows that
\begin{equation}
E_a(s) = -\lim_{N\to 0}\frac{\partial_d E(s)}{4\pi N (R+d)^2}.
\end{equation}
By virtue of the fact that the Casimir energy is zero for an atom in vacuum
($Q = 0$) without boundaries, we define the interaction energy by the following
relation
\begin{equation}
E_{\Omega} = \lim_{s\to 0} \{E_a(s) -  \lim_{\Omega \to 0} E_a(s)\}.
\end{equation}
With this definition we integrate by part over $k$ and arrive with the final
formula ($x=kR,\ z=kL,\ L = R + d$)
\begin{equation}
E_\Omega = -\frac{\hbar c\Omega}{\pi L^2}\! \sum_{l=1}^\infty \nu \!\!\!
\int\limits_0^\infty\!\! dk \alpha (i\omega) \left\{ \frac{s_l^2(x)
e_l^2(z)}{f_{\te}(ik)} + \frac{{s'}_l^2(x) {e'}_l^2(z) + {s'}_l^2(x){e}_l^2(z)
\frac{\nu^2 - \frac{1}{4}}{z^2}}{f_{\tm}(ik)} \right\},\label{Fcp}
\end{equation}
where the Jost functions on the imaginary axes read
\begin{eqnarray}
f_{\te}(ik) &=& 1 + \frac{\Omega}{k} s_l(x)e_l(x) ,\\
f_{\tm}(ik) &=& 1 - \frac{\Omega}{k} s'_l(x)e'_l(x).
\end{eqnarray}

To perform computations one needs an expression for the atomic dynamic
polarizabilities of hydrogen. It was shown\cite{BlaKliMos05} that
the polarizabilities can be represented with sufficient precision in the
framework of the single-oscillator model
\begin{equation}
 \alpha (i\omega) = \frac{g_a^2}{\omega^2 + \omega^2_a},\label{alpha}
\end{equation}
where $\alpha_a (0) = 4.50\ a.u.$ ($1\ a.u. = 1.482 \cdot 10^{-31} m^3$) and
$\omega_a = 11.65 eV$ for hydrogen atom.

Let us consider different limits.

1) In the limit of perfect conductivity, $\Omega\to\infty$, which we call the
Boyer limit, we obtain
\begin{equation}
E_B = -\frac{\hbar c}{\pi L^2}\!\sum_{l=1}^\infty\! \nu \!\!\int_0^\infty
\!\!\!\!\!\!dk k \alpha (i\omega)
\!\!\left\{ \frac{s_l^2(x) e_l^2(z)}{s_l(x)e_l(x)} - \frac{
{s'}_l^2(x) {e'}_l^2(z) + {s'}_l^2(x)
{e}_l^2(z) \frac{\nu^2 - \frac{1}{4}}{z^2}}{s'_l(x)e'_l(x)}
\right\}.\label{FcpBoyer}
\end{equation}

2) The limit of infinite radius of sphere, $R\to \infty$, with fixed distance,
$d$, between the surface of sphere and an atom requires more machinery. In
this case we change the variable of integration $k\to \nu k$  in Eqs.
(\ref{Fcp}) and (\ref{FcpBoyer}) and use the uniform expansion for Bessel
functions.\cite{AbrSte70} In the limit of $R\to\infty$, the integrands in
above both  expressions have the same form and the main contribution to the
energy comes from the first term of uniform expansion,
 \begin{equation}
E = -\lim_{R\to \infty} \frac{\hbar c g^2}{\pi c^2 (R+d)^2} \sum_{l=1}^\infty
\nu^3 \int_0^\infty \frac{dy y}{y^2 \nu^2 + q^2}\frac{e^{-2\nu [\eta(u) -
\eta(y)]}}{ut(u)},
\end{equation}
where $u = y (1+d/R),\ q_a = k_a R$ , $t(x) =
1/\sqrt{1+x^2}$ and $\eta (x) = \sqrt{1+x^2} + \ln \frac{x}{1+\sqrt{1+x^2}}$.

\begin{figure}[pb]
%\centerline{\psfig{width=6cm,file=1.ps}}
\centerline{\includegraphics[scale=0.9]{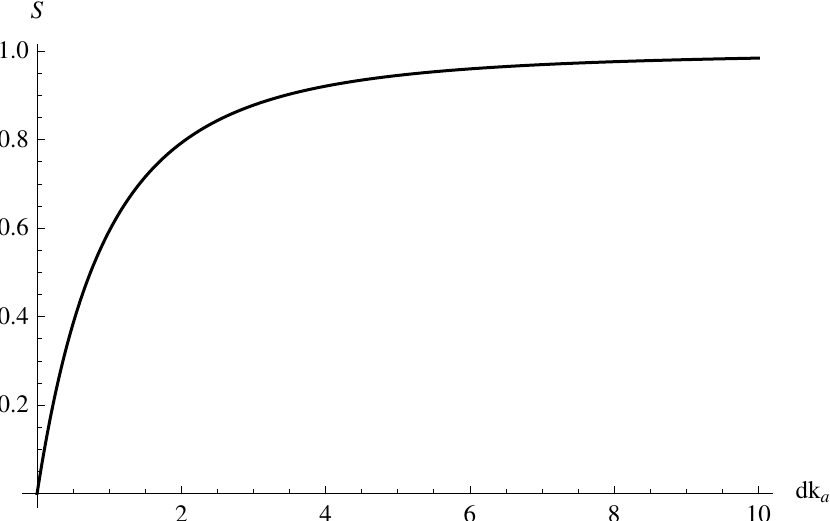}}
\vspace*{8pt}
\caption{\footnotesize The plot of $S$ as the function of the $v=kd_a$. It tends
to unity for large $v$ ($E \sim d^{-4}$) and it is linear over $v$ ($E\sim
d^{-3}$) for small distances between an atom and plate. The relation of the
energy and $S$ is given by Eq. (\ref{defS}).}\label{fig:sboyerv}
\end{figure}

Next, the sum over $l$ we represent in the following integral
\begin{equation}
\sum_{l=1}^\infty \frac{\nu^3 e^{-2\nu\delta}}{y^2 \nu^2 + q_a^2} = \frac{1}{4
q_a y} \int_{0}^\infty \frac{27 + 17 e^{-2(t+\delta)} + 5
e^{-4(t+\delta)}-e^{-6(t+\delta)}} {e^{3(t+\delta)}(e^{-2(t+\delta)}-1)^4 }
\sin\frac{2q_at}{y} dt.
\end{equation}
Assuming this expression we interchange the limit $R\to \infty$ and integrals
over $y$ and $t$ and obtain
\begin{equation}\label{defS}
E= -\frac{3\hbar c \alpha(0)}{8\pi d^4} S,
\end{equation}
where
\begin{equation}
S = \frac 13 \int_0^\infty dt e^{-t} \left\{ \frac{1+t}{1+\frac{t^2}{4v^2}} +
\frac{t}{(1+\frac{t^2}{4v^2})^2} \right\},
\end{equation}
and $v = dk_a$. Let us consider large distance, $d$, between the plate (sphere
of infinite radius) and an atom, $d k_a \gg 1$. In the limit of $v\to\infty$ we
obtain that $S =1$ and therefore the Casimir-Polder $(\sim d^{-4})$ energy,
\begin{equation}
E= -\frac{3\hbar c \alpha(0)}{8\pi d^4},
\end{equation}
is recovered. For small distances, $dk_a \ll 1$, we change the variable $t\to
\tau = t/2v$ and take the limit of $v \to 0$. In this case we obtain that $S =
\pi v/3$ and the energy has the form $\sim d^{-3}$,
\begin{equation}\label{CasPolSmall}
E= -\frac{\hbar c \alpha(0)k_a}{8 d^3},
\end{equation}
as should be the case. The plot of the $S$ as function of variable $v=dk_a$ is
shown in Fig. \ref{fig:sboyerv}.

3) Let us analyze the energy for large ($d\gg k_a^{-1}, d\gg R$) and small
($d\ll k_a^{-1}, d\ll R$) distances between the sphere and an atom for finite
$\Omega$ and $R$. In the case of large distance, $d\to \infty$, of an atom from
the shell we use  Eq. (\ref{Fcp}). We change integrand variable $k=y/d$, next
take limit $d\to \infty$, and then we take the integral over $y$. The main
contribution comes from the first term with $l=1$:
\bs
\begin{eqnarray}
E_\Omega &\approx&  -\frac{3\hbar c \alpha(0)}{8\pi d^4} S_\Omega,\adb \\
S_\Omega &=& \frac{R^{3}}{d^{3}} \left\{ \frac{7Q}{3(3+Q)} + \frac{46}{3}
F(a)\right\},\adb \label{sdinf} \\
F(a) &=& \frac{8a^{2}}{23} \int_0^\infty \frac{y^{4} + 2 y^{3} + 5 y^{2} + 6 y +
3}{3y^{2} + 2a^{2}}e^{-2y}dy,
\end{eqnarray}
\es
where $a^{2} = Q d^{2}/R^{2} = d^{2} \Omega /R$. The first term in above
expression (\ref{sdinf}) comes from \TE\ mode and second -- from \TM\
polarization. The function $F$  increases monotonically from zero for small $a$
($d^{2} \ll R/\Omega$) to unity  for large $a$ ($d^{2} \gg R/\Omega$). In the
case of $a \ll 1$ the function $F(a) \approx 2\pi \sqrt{6} a/23$. Therefore, in
the limit of $\Omega \to 0$, the energy $E_\Omega \to 0$ as should be the case.

Assuming a finite conductivity, $\Omega \not = 0$, and  large distance $d \gg
k_a^{-1}, d \gg R, d \gg \sqrt{R/\Omega}$  we obtain that
\begin{equation*}
S_\Omega = \frac{R^3}{d^3} \left\{ \frac{7Q}{3(3+Q)} + \frac{46}3 \right\}
\end{equation*}
and  we arrive with expression
\begin{equation}\label{Egreat}
E_\Omega \approx  -\frac{\hbar c \alpha (0) R^3}{8\pi  (3+Q)d^7} (53 Q + 138).
\end{equation}
Taking into account the Casimir-Polder interaction energy of two atoms with
polarizations $\alpha$ and $\alpha_f$,
\begin{equation}
E = - \frac{23}{4\pi} \frac{\hbar c \alpha(0) \alpha_f(0)}{d^7},
\end{equation}
we observe that the sphere with finite conductivity has static polarizability
\begin{equation}\label{alpha_2}
\alpha_f = \frac{53 Q + 138}{46 Q + 138}R^3.
\end{equation}

For small distances we obtain that
\begin{equation}
E= -\frac{\hbar c \alpha(0)k_a}{8 d^3}
\end{equation}
as should be the case, because close to the sphere we observe flat surface.

\section{Numerical Analysis}\label{Sec:Num}

For simplicity we extract as a factor the Casimir-Polder expression for the
interaction energy of an atom with plate,
\begin{equation}
 E_{\Omega,B} = -\frac{3\hbar c \alpha (0)}{8\pi d^4} {S}_{\Omega},
\end{equation}
and we will numerically calculate the dimensionless quantity ${S}_\Omega$.

Let us consider the interaction energy between hydrogen atom and molecule
$C_{60}$. For this molecule\cite{Bar04} we have: $R = 3.42 \textrm{\AA} = 0.342
nm$, $Q = \Omega R = 4.94 \cdot 10^{-4}$ and $\Omega/k_a = 2.44\cdot 10^{-2}$.
The polarizability of hydrogen atom within the single-oscillator model
reads\cite{RauKleColBru82,BorGeyKliMos06,BlaKliMos07} $\alpha_a (0) = 4.50\ a.u.$
($1\ a.u. = 1.482 \cdot 10^{-31} m^3$) and $\omega_a = 11.65 eV = 17.698 \cdot
10^{15} Hz\ (k_a = 0.059 nm^{-1},\ \lambda_a = 106.4 nm)$ where $\omega/c = k=
2\pi /\lambda$. Therefore, $q_a = k_a R = 0.0202$.

\begin{figure}[pb]
\centerline{\includegraphics[scale=0.7]{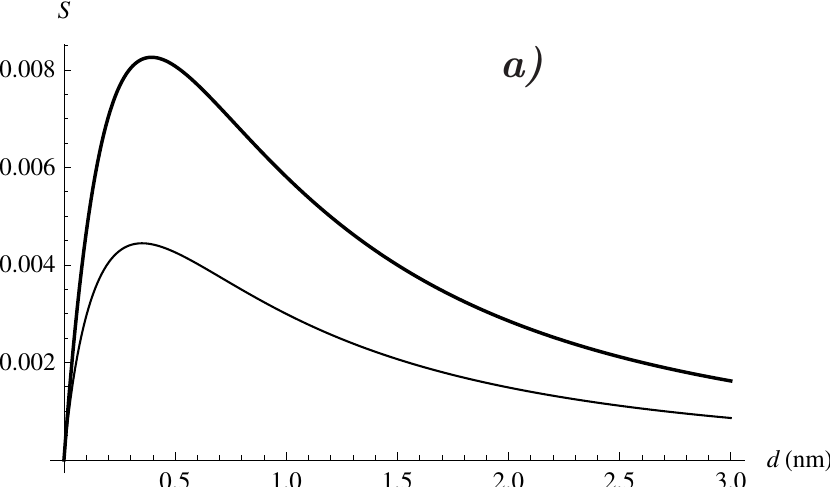}\includegraphics[scale=0.7]{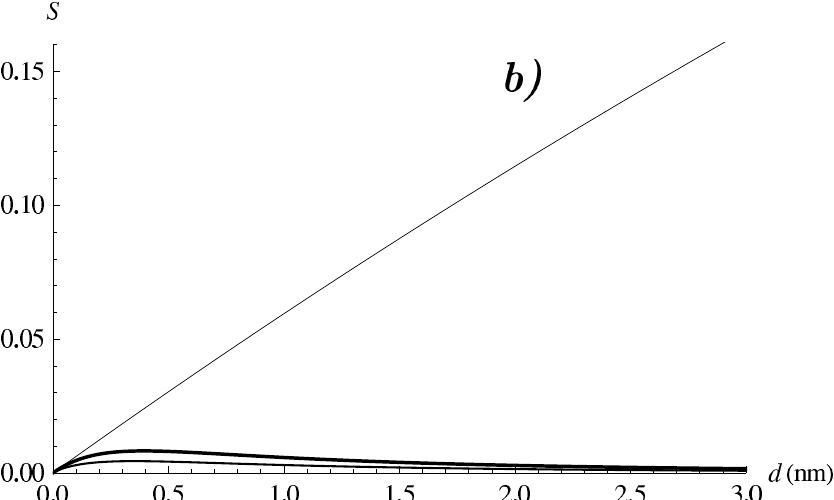}}
\vspace*{8pt}
\caption{\footnotesize The plot of $S$ as the function of the distance $d$
between an atom and the sphere. Thin curve is the energy for the case $R \to
\infty$ (Casimir-Polder energy for plate), middle thickness curve is the case of
the molecule $C_{60}$, and the thick curve is the case of ideal sphere ($\Omega
\to \infty$). In the figure $b$ we compare the energy for the plane with the
energy in the sphere case.}\label{fig:srd}
\end{figure}

\begin{figure}[pb]
\centerline{\includegraphics[scale=0.7]{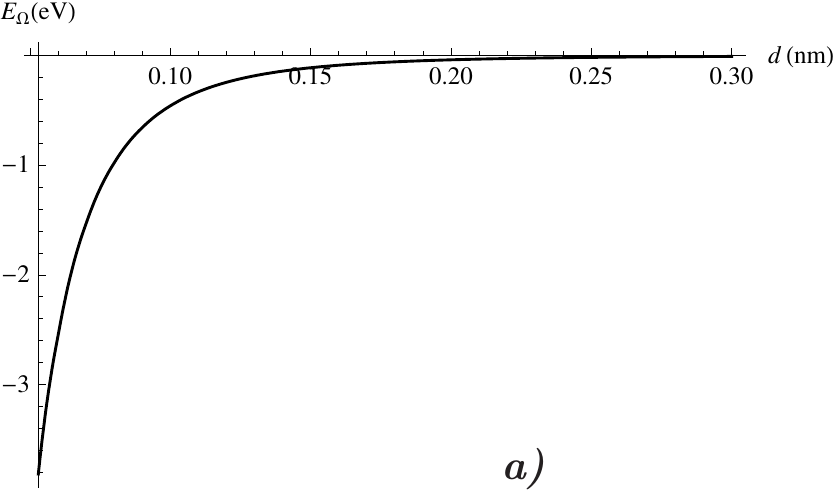}\includegraphics[scale=0.7]{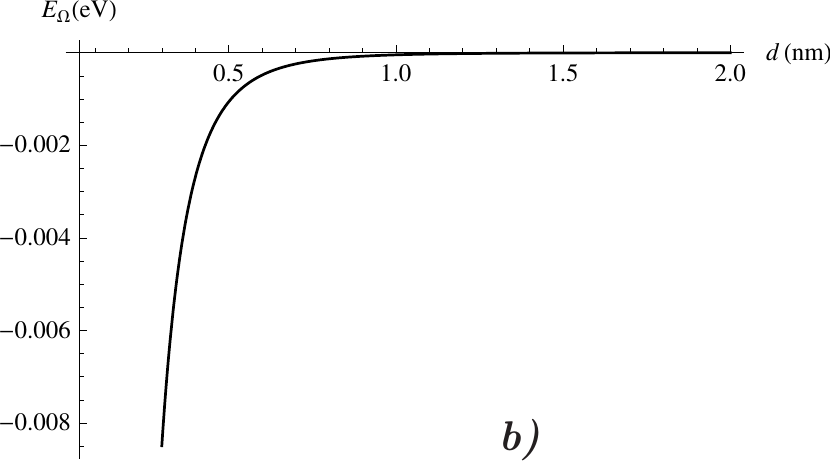}}
\vspace*{8pt}
\caption{\footnotesize The plot of the energy $E_\Omega$ as the function of the
distance $d$ between the sphere and the hydrogen atom. In the figure $a)$ we
show the energy starting from the distance $d = 0.053 (nm)$ (the radius of the
hydrogen atom). In the  figure $b)$  the energy in large interval is shown.}
\label{fig:srdhyd}
\end{figure}

Taking into consideration  all the numerical values of parameters we represent
the energy for this system in the following form
\begin{equation}
 E_{\Omega}(eV) = -\frac{0.0156}{d^4(nm)} S_{\Omega}(q_a,r),
\end{equation}
where the energy is measured in $eV$ and the distance is measured in nanometres.
The numerical simulations for the function $S$ are shown in Fig. \ref{fig:srd}
and the energy $E_\Omega$ in Fig. \ref{fig:srdhyd}.
The radius of the hydrogen atom is $r_H = 0.053 nm$. For this minimal distance,
$d=r_H$, we have numerically $E = 3.8 eV$. In the case of plate with hydrogen
atom we obtain $6.4 eV$.

For large distances we obtain from Eq. (\ref{Egreat})
\begin{equation}\label{Efar}
E_\Omega(eV) \approx -\frac{0.0095}{d^7(nm)}.
\end{equation}
This expression approximates the exact one with error $10\%$ starting with
distance $d=50nm$.

\section{Conclusion} \label{Sec:Conc}

In the foregoing, we have obtained the analytic expression for the
Casimir-Polder (van der Waals) energy for a system which contains an atom or
microparticle and  infinitely thin sphere with finite conductivity which models
a fullerene. We used the zeta-regularization approach and for renormalization we
used a simple physically reasonable condition -- the energy should be zero for
an  atom alone without a sphere. The conductive sphere with radius $R$ is
characterized by the only parameter $\Omega = 4\pi n e^2/mc^2$ with dimension of
wave number, where $n$ is the surface density of electrons. The limit $\Omega
\to \infty$ corresponds to the ideal case considered by Boyer.\cite{Boy68} The
microparticle is characterized by the only parameter, polarizability $\alpha$.

The expression obtained reproduces in the limit $R\to \infty$ the Casimir-Polder
result for an atom and plate (see Eqs. (\ref{defS})-(\ref{CasPolSmall})). For
small distances we have $d^{-3}$ dependence and far from the plate we obtain
$d^{-4}$ due to retardation. For finite radius of the sphere we have different
behavior of the energy. Close to the sphere,  $d\ll 1/k_a$ and $d\ll R$, we have
the same $d^{-3}$ dependence as in the Casimir-Polder case and far from the
sphere we obtained $d^{-7}$ dependence given in Eq. (\ref{Egreat}). This
expression is valid for $d\gg 1/k_a$ and $d\gg R$.

Application to the molecule $C_{60}$ with hydrogen atom is plotted in Fig.
\ref{fig:srdhyd}. For closest distance atom from the fullerene, which is radius
of hydrogen atom $r_H$, the energy is $3.8 eV$ which is two times smaller then
for the case of hydrogen atom with plate. Away from the fullerene (in fact
larger then $50 nm$) the energy falls down as $d^{-7}$ (see Eq. (\ref{Efar}))
which is in three orders of
magnitude faster then for the Casimir-Polder case. This dependence corresponds
to the Casimir-Polder interaction atoms for large distance. Taking into account
this analogy we obtain the polarizability of fullerene ($Q = \Omega R =
4.94\cdot 10^{-4} \ll 1$)
\begin{equation*}
\alpha_f = \frac{53 Q + 138}{46 Q +138} R^3 \approx R^3 = 4\cdot 10^{-29} m^3.
\end{equation*}
This expression is close to that calculated in Ref. \refcite{FowLazZan91} where
the authors obtained $\alpha_p(0) = 7\cdot 10^{-29} m^3$.

\section*{Acknowledgments}
The author would like to thank V. Mostepanenko and G. Klimchitskaya for
stimulation of
these calculations and M. Bordag for discussions. This work was supported by the
Russian Foundation for Basic Research Grant No. 11-02-01162-a.

\end{document}